\documentclass[aps,prl,reprint,groupedaddress,showkeys]{revtex4-1}

\usepackage[utf8]{inputenc}

\usepackage{fullpage}
\usepackage[T1]{fontenc}
\usepackage{verbatim}
\usepackage{amssymb,amsmath}
\usepackage{xcolor}
\usepackage[colorlinks,linktocpage,linkcolor=blue,citecolor=red,urlcolor=red]{hyperref}

\def\+{{+\!\!\!+}}

\def\be{\begin{equation}} 
\def\ee{\end{equation}}

\def\p{\partial}

\newcommand{\D}{\mathrm{D}}
\newcommand{\Db}{\overline{\mathrm{D}}}

\allowdisplaybreaks

\begin{document}

\title{The 2D Volkov--Akulov model as a \texorpdfstring{$T \overline{T}$}{TTbar} deformation}
\author{Niccol\`o Cribiori}
\email{niccolo.cribiori@tuwien.ac.at}
\affiliation{Institute for Theoretical Physics, TU Wien, Wiedner Hauptstrasse 8-10/136, A-1040 Vienna, Austria}
\author{Fotis Farakos}
\email{fotios.farakos@kuleuven.be}
\affiliation{KU Leuven, Institute for Theoretical Physics, Celestijnenlaan 200D, B-3001 Leuven, Belgium}
\author{Rikard von Unge}
\email{unge@physics.muni.cz}
\affiliation{Institute for Theoretical Physics, Masaryk University, 611 37 Brno, Czech Republic}

\begin{abstract}
We show that the two-dimensional $N=(2,2)$ Volkov--Akulov action that describes the spontaneous breaking of supersymmetry is a $T\overline{T}$ deformation of a free fermionic theory. Our findings point toward a possible relation between nonlinear supersymmetry and $T \overline T$ flows.

\end{abstract}

\maketitle

\section{I. Introduction} 

An interesting approach in the study of quantum field theories consists in starting with an accessible model, for example a free theory, and perturbing it by means of some deformation, which induces a renormalization group flow.
Mostly one is interested in relevant deformations, which keep the theory well defined in the UV, whereas irrelevant deformations change the very definition of the theory in an uncontrollable way. An exception to this statement is given by a special class of irrelevant deformations of two-dimensional theories, namely $T\overline T$ deformations ($T$ being the energy-momentum tensor), since they happen to preserve some properties of the original theory along the flow \cite{Zamolodchikov:2004ce, Smirnov:2016lqw, Cavaglia:2016oda}. For example, the finite volume spectrum of the deformed theory can be determined from that of the undeformed one \cite{Zamolodchikov:2004ce,Smirnov:2016lqw}. In this Letter we focus on the relation between $T \overline T$ deformations and supersymmetry, which was recently investigated in \cite{Baggio:2018rpv,Chang:2018dge,Jiang:2019hux,Chang:2019kiu,Coleman:2019dvf}. 

Considering a two-dimensional quantum field theory, its $T\overline T$ deformation can be defined by the flow equation 
\be
\label{TTdefI}
\partial_\lambda {\cal L} = - \det(T[{\cal L}_\lambda]) \, , 
\ee
where $T[{\cal L}_\lambda]$ is the energy-momentum tensor of the deformed Lagrangian, 
which in light-cone coordinates can be expressed as
$\det(T) = T_{\+\+} T_{==} - \Theta^2$, where $\Theta = T_{\+=}= T_{=\+}$. 
It is important to observe that \eqref{TTdefI} holds with the use of the equations of motion, since $T$ is defined up to terms that vanish on-shell. 

Equation \eqref{TTdefI} is nonlinear and in principle hard to solve. However, in some cases a one parameter family of Lagrangians solving it is explicitly known. 
For example, in \cite{Cavaglia:2016oda,Bonelli:2018kik}, it is shown that, starting with a free scalar as initial data
\be
{\cal L}_0 = \frac12 \p_\+ \phi \p_= \phi \, , 
\ee
the $T\overline T$ flow equation can be solved recursively and the result is the Nambu--Goto Lagrangian
\be
{\cal L}_{NG} = \frac{1}{2 \lambda} \left( -1 + \sqrt{1 + 2 \lambda \p_\+ \phi \p_= \phi } \right) \, . 
\ee

We wonder if an equivalent procedure can be followed for a purely fermionic theory. Our starting point is a free Lagrangian for a pair of 2D complex fermions $(G_+,G_-)$ \footnote{Through this letter we follow the conventions of \cite{Farakos:2016zam}.} 
\be
{\cal L}_0 = i G_-\p_{\+}\overline G_- + i G_+ \p_=\overline G_+ \, . 
\ee
The nonvanishing components of the energy-momentum tensor are then
\be
\begin{aligned}
T_{\+\+} &  \sim i \overline G_+ \p_\+ G_+ 
+ i G_+ \p_\+ \overline G_+ \, , 
\\
T_{==}& \sim i \overline G_- \p_= G_- 
+ i G_- \p_= \overline G_- \, , 
\end{aligned}
\ee 
and the equations of motion read $\partial_= G_+  = \partial_\+ G_-=0$.
When trying to solve \eqref{TTdefI} recursively, at the first step, one finds that
\be
\det(T[{\cal L}_0]) = T_{\+\+} T_{==} 
\sim \overline G_+ \overline G_- \square  (G_- G_+) \, ,  
\ee
using also the equations of motion. We notice therefore that, at the first order in the deformation of the free fermionic theory, a term is produced which has precisely the form of the four-fermi term in the Volkov--Akulov model \cite{Volkov:1973ix}. This gives rise to the question of whether or not the Volkov--Akulov model is a solution of the $T\overline T$ flow equation, such that in the limit $\lambda=0$ the free theory is recovered. In the rest of this Letter we will verify this expectation. In particular, we will show that, for the two-dimensional $N=(2,2)$ Volkov--Akulov model, 
\be
\label{VAcomp}
\begin{aligned}
{\cal L}_{VA} &= -f^2 + i G_-\p_{\+}\overline G_- + i G_+ \p_=\overline G_+ \\
&- \frac{1}{f^2}G_+G_-\square(\overline G_- \overline G _+)\\
&-\frac{1}{f^6}G_+G_-\overline G_-\overline G_+\square(G_+G_-)\square(\overline G_- \overline G_+) \, ,  
\end{aligned}
\ee
formula \eqref{TTdefI} holds, with $\lambda$ related to the supersymmetry breaking scale by $2 \lambda^{-1} = f^2$. 

It is worth mentioning that a possible relation between $T\overline T$ deformations and the Volkov--Akulov model was already suggested several years ago \cite{Kastor:1988ef,Zamolodchikov:1991vx}, but a clear answer on whether or not such a proposal was correct has never been given. Instead, it has been argued more recently \cite{Bonelli:2018kik} that the $T\overline T$ deformation of a Goldstino might contain infinite interactions. This Letter aims to address precisely this long-standing open problem, showing explicitly that the $T\overline T$ deformation of a free fermion gives raise to the Volkov--Akulov model, whose Lagrangian can be written in a closed form.

\section{II. The supercurrents of the 2D Volkov--Akulov model}

To facilitate the calculations, we use $N=(2,2)$ superspace. The nonvanishing superspace anticommutators describing the two-dimensional $N=(2,2)$ supersymmetry algebra without central charges are
\begin{equation}
\begin{aligned}
\{ \D_- , \Db_- \} = i \p_{=} \, , \quad
\{ \D_+ , \Db_+ \} = i \p_{\+} \, . 
\end{aligned} 
\end{equation} 
Consider a set of $N=(2,2)$ chiral superfields $\Phi^i$, defined by $\Db_{\pm} \Phi^i = 0$. The most general two-derivatives supersymmetric $\sigma$-model Lagrangian describing their interactions has the form
\be
\label{Lgen}
{\cal L} = \int d^4 \theta K(\Phi^i, \overline \Phi^j) + \left( \int d^2 \theta W(\Phi^i) + c.c. \right) \, , 
\ee
where the K\"ahler potential $K$ is a real function of the $\Phi^i$, while the superpotential $W$ is holomorphic. To simplify the derivation of our result, we can also restrict the analysis to the case in which the K\"ahler manifold is flat, therefore $K = \Phi^i \overline \Phi_i$, where $\overline \Phi_i =\delta_{ij} \overline \Phi^j$. The equations of motion in superspace which stem from \eqref{Lgen} read 
\be
\label{SEOM}
\Db_- \Db_+ \overline \Phi_i = - \partial_i W \, , 
\ee
where we define $\partial_i W = \partial W / \partial \Phi^i$. 
By expanding \eqref{SEOM} in the chiral $\theta$ coordinates, it is possible to extract the complete set of equations of motion for the component fields. We stress that \eqref{SEOM} hold for any choice of $W$ and for any set of chiral superfields $\Phi^i$. 

Along with the proposal of \cite{Chang:2019kiu}, the $N=(2,2)$ supersymmetric extension of the $T\overline T$ deformation is given (roughly) by the square of the supercurrent multiplet. Indeed, this multiplet can be described by an $N=(2,2)$ superfield that contains in its component expansion different Noether (conserved) currents, such as the supercurrent itself and the energy-momentum tensor. The complete set of conservation equations for these component Noether currents can be embedded into one single superspace equation. For the model \eqref{Lgen} that we are considering, the explicit form of the two-dimensional version of the Ferrara--Zumino multiplet is given by \cite{Chang:2019kiu,Ferrara:1974pz}
\be
\label{JJ}
\begin{aligned}
J_\+ & =  \frac12 [\D_+ , \Db_+] K 
+ i \frac{\partial K}{\partial \Phi^i} \partial_\+ \Phi^i 
+ c.c. 
\\ 
J_= & =  \frac12 [\D_- , \Db_-] K 
+ i \frac{\partial K}{\partial \Phi^i} \partial_= \Phi^i +c.c.
\end{aligned}
\ee
and its conservation equation reads 
\be
\label{SCOM}
\Db_+ J_{=} = - \D_- Z \, , \quad  \Db_- J_{\+} = \D_+ Z \, , 
\ee
where $Z$ is a chiral $N=(2,2)$ superfield defined as \cite{Chang:2019kiu,Ferrara:1974pz} 
\be
\label{Z}
\begin{aligned}
Z  & = 2 W(\Phi^i) \, . 
\end{aligned}
\ee
As a check, by using the superspace equations of motion \eqref{SEOM} one can prove that  \eqref{SCOM} holds. 

Generically, the Noether currents are defined up to improvement terms. The procedure we follow is to automatically introduce such terms in order to make the currents compatible with supersymmetry. For example, when considering the energy-momentum tensor on a flat background, one can always add an improvement term proportional to the background metric, without spoiling the conservation equation. Such an improvement term is indeed important in the evaluation of the energy-momentum tensor of the Volkov--Akulov model. In the supersymmetric procedure presented below, it is automatically taken into account.

Equivalently to the four-dimensional constructions \cite{Rocek:1978nb,Casalbuoni:1988xh}, a 2D Goldstino can be described by an $N=(2,2)$ chiral superfield $X$ that satisfies the additional nilpotency constraint 
\be
\label{X2zero}
X^2 = 0 \, . 
\ee
This constraint admits a solution describing the complete spontaneous breaking of $N=(2,2)$ supersymmetry. Such a solution is a chiral superfield with expansion \cite{Farakos:2016zam} 
\be
X = \frac{G_- G_+}{F} 
+ \theta_+ G_- 
+ \theta_- G_+ 
+ \theta_+ \theta_- F 
\, , 
\ee
where $G_+$ and $G_-$ are the two Goldstini, while the field $F$ is auxiliary and it acquires a nonvanishing vacuum expectation value.

The constraint \eqref{X2zero} is implementing a nonlinear realization of supersymmetry. For future purposes, however, it is convenient to maintain supersymmetry linearly realized off-shell. This can be done by introducing the constraint \eqref{X2zero} at the Lagrangian level, by means of a chiral Lagrange multiplier superfield $M$. In view of the generic model \eqref{Lgen}, we consider therefore a set
\be
\label{setchiralSF}
\Phi^i=\{X,M\},
\ee
governed by a K\"ahler potential $K = X \overline X$ and a superpotential $W=f X + M X^2$, where $f$ is set to be real. The resulting Lagrangian is then 
\be
\label{NILP}
{\cal L} = \int d^4 \theta X \overline X 
+ \left( \int d^2 \theta ( f X + M X^2 ) + c.c. \right) \, .  
\ee
We recall that, even if $M$ does not appear in $K$, the previous discussion on the supercurrents and the superspace equations of motion still holds for the set \eqref{setchiralSF}. By varying \eqref{NILP} with respect to $M$, we recover the constraint \eqref{X2zero}, and the Lagrangian takes the form 
\be
\label{NILP2}
{\cal L} = \int d^4 \theta X \overline X 
+ \left( f \int d^2 \theta X  + c.c. \right) \, . 
\ee
Expanding it in components and after integrating out $F$, the Lagrangian reduces to \eqref{VAcomp}. This verifies that \eqref{VAcomp} is the two-dimensional Volkov--Akulov model describing the interactions of two Goldstini.

The complete set of superspace equations of motion can be obtained by varying \eqref{NILP} with respect to both $X$ and $M$. They read
\be
\label{LMEOM}
\begin{aligned}
\Db_- \Db_+  \overline X  = - f - 2 M X \, , \qquad 
X^2  = 0 \, . 
\end{aligned}
\ee
From these equations we can also derive the constraint 
\be
\label{R2}
X \Db_- \Db_+  \overline X  = - f X \,, 
\ee 
which was proposed together with \eqref{X2zero} in \cite{Rocek:1978nb}. It corresponds to eliminating only the auxiliary field $F$ of the nilpotent chiral superfield $X$, but it does not imply any other equation of motion. In particular, even when imposing \eqref{R2}, the fermions of the theory are still completely off-shell and therefore it is possible to insert such a constraint back into the Lagrangian, without creating inconsistencies. Imposing both \eqref{X2zero} and \eqref{R2} on the Lagrangian \eqref{NILP}, we obtain then 
\be
\label{VAR}
{\cal L} = f \int d^2 \theta X \, , 
\ee
which is real up to boundary terms. This Lagrangian \eqref{VAR} is yet another way to write the Volkov--Akulov model in superspace.

We now have all of the ingredients to evaluate the supercurrent superfields for the Volkov--Akulov model. Using the definitions of $J_\+$, $J_=$ and $Z$ from \eqref{JJ} and \eqref{Z}, the K\"ahler potential and superpotential of \eqref{NILP}, together with \eqref{X2zero} and \eqref{R2}, we find that 
\be
\label{ZVA}
Z = 2 f X \, , 
\ee
and 
\be 
\label{JVA}
J_\+ = 2 \D_+ X \Db_+ \overline X \, ,  \quad 
J_= = 2 \D_- X \Db_- \overline X \, . 
\ee
When checking that these superfields satisfy \eqref{SCOM} once the superspace equations of motion are used, the following equations stemming from \eqref{LMEOM} can be helpful
\be
\D_\pm X \Db_- \Db_+  \overline X  = - f \D_\pm X \, . 
\ee 
Finally, notice that the Lagrange multiplier superfield $M$ has dropped out of the supercurrent superfields.

\section{III. The TT-bar deformation}

For an $N=(2,2)$ supersymmetric theory, the authors of \cite{Chang:2019kiu} propose the following form for the $T\bar T$ deformation 
\be
\label{TTdef}
\det(T[{\cal L}]) = \frac18 \int d^4 \theta \left( J_\+ J_= - 2 Z \overline Z \right) \, .  
\ee 
We now investigate this proposal in the case of the Volkov--Akulov model. 
From \eqref{ZVA} and \eqref{JVA}, we directly find that
\be
J_\+ J_= - 2 Z \overline Z = - 4 f^2 X \overline X \,  
\ee
and as a result, we have 
\be 
\label{TLone}
\det(T[{\cal L}]) = \frac{f^2}{2} \int d^4 \theta  X \overline X 
= \frac{f^3}{2} \int d^2 \theta X  \, .  
\ee
To study the flow equation, we can identify the supersymmetry breaking scale with $\lambda$ as in \cite{Jiang:2019hux}. This means that
\be
\label{FL}
f^2 = 2 \lambda^{-1} \, , \quad \lambda>0 \, . 
\ee 
We then vary the Volkov--Akulov action with respect to $f$. In particular, we can use the off-shell superspace form \eqref{NILP2} that gives 
\be
\frac{\partial {\cal L}}{\partial f} = \int d^2 \theta X  + \int d^2 \overline \theta \overline X \, .  
\ee
Finally, with the use of the superspace constraint \eqref{R2} and the relation \eqref{FL}, we find that 
\be
\frac{\partial {\cal L}}{\partial \lambda} = -\frac{f^3}{2} \int d^2 \theta X = - \det(T[{\cal L}]) \, .  
\ee
The flow equation \eqref{TTdefI} is therefore verified for the Volkov--Akulov model. 
Since the deformation parameter $\lambda$ is related to the scale of supersymmetry breaking through $f^2 = \frac{2}{\lambda}$, the limit of vanishing deformation, namely $\lambda\rightarrow 0$, corresponds to sending the scale of supersymmetry breaking to infinity. In this case the Volkov--Akulov theory \eqref{VAcomp} reduces to a pair of free fermions \footnote{In the limit $\lambda\to0$, the constant term in the Volkov--Akulov action diverges. As long as we are in flat space, this (infinite) constant can be ignored.}. We then notice an interesting analogy with the case of the free scalar presented in the Introduction.

\section{IV. Discussion}

In this letter we showed that the 2D Volkov--Akulov model is a solution to the $T \overline T$ flow equation. We adopted a manifestly supersymmetric approach, following the proposal of \cite{Chang:2019kiu}, but one can verify our result independently, by inserting \eqref{VAcomp} directly into \eqref{TTdefI}. We also performed this calculation as a consistency check. We recall that the Volkov--Akulov model is universal, namely it describes an indispensable part of the low energy spectrum of any model with spontaneous supersymmetry breaking. 
Therefore our findings apply to a very large class of theories.

To the best of our knowledge, we have presented the first example of a purely fermionic construction that solves the flow equation \eqref{TTdefI} in a closed form and that has manifest nonlinearly realized $\mathcal{N}=(2,2)$ supersymmetry. Indeed, in other constructions, as in \cite{Bonelli:2018kik}, there is no supersymmetry present in the pure fermionic models and no hint that the $T\overline T$ deformation of the free fermion could have a nonlinearly realized supersymmetry. On the other hand, in \cite{Chang:2019kiu}, it was suggested that the pure fermionic sectors of their constructions should themselves be $T\overline T$ deformations, however, these fermionic Lagrangians were not explicitly presented. Similarly, more recently in \cite{Frolov:2019nrr}, a large class of models were studied that included fermionic systems, but again there was no indication of nonlinear supersymmetry.

Our findings suggest a connection between $T\overline T$ deformations and nonlinear supersymmetry that would be worth investigating for matter-coupled Volkov--Akulov models as well. Other systems with $N=(2,2)$ supersymmetry were recently proposed in \cite{Chang:2019kiu} and hint at a relation between (partial) supersymmetry breaking and $T \overline T$ deformations. This direction deserves further study. It is also worth investigating different types of supersymmetry breaking models, for example those in \cite{Farakos:2016zam}, and understanding which of these Lagrangians can be interpreted as $T\overline T$ deformations. Given that 2D $\mathcal{N}=(2,2)$ supersymmetry corresponds to 4D $\mathcal{N}=1$, a generalization of our result to higher dimensions is also compelling. Recalling that the 4D Volkov--Akulov model describes the Goldstino on an anti-D3-brane \cite{Kallosh:2014wsa}, one might be able to give an interpretation of $T\overline T$ deformations in terms of extended objects in string theory. On a similar footing, the structure of Lagrangians with nonlinearly realized supersymmetry may also pave the way to study the $T\overline T$ deformations in higher dimensions \cite{Bonelli:2018kik,Chang:2018dge,Taylor:2018xcy,Conti:2018jho}.

Finally, let us note that one of the most interesting aspects of the $T\overline T$ deformation is that, classically, it gives rise to the Nambu--Goto action as a deformation of the free scalar. Similarly, the Born--Infeld action that relates to effective actions for D-branes arises as a $T\overline T$ deformation of the free Maxwell theory \cite{Conti:2018jho,Chang:2018dge}. It is therefore gratifying to see that the Volkov--Akulov model finds its place within these fundamental actions.

\begin{acknowledgments}
 The work of NC is supported by FWF Grant No. P 30265. 
 The work of FF is supported by KU Leuven C1 Grant No. ZKD1118 C16/16/005. 
 FF thanks the theory group of MU Brno for support and hospitality during the early stages of this work. NC and FF thank the organizers of Pre-strings 2019 and Strings 2019, where part of this work was carried out. 
\end{acknowledgments}



\begin{thebibliography}{30}

\bibitem{Zamolodchikov:2004ce}
A.~B. Zamolodchikov, \emph{{Expectation value of composite field T anti-T in
  two-dimensional quantum field theory}},
\href{http://arxiv.org/abs/hep-th/0401146}{{\tt arXiv:hep-th/0401146
  [hep-th]}}.

\bibitem{Smirnov:2016lqw}
F.~A. Smirnov and A.~B. Zamolodchikov, \emph{{On space of integrable quantum
  field theories}},
  \href{http://dx.doi.org/10.1016/j.nuclphysb.2016.12.014}{Nucl. Phys. {\bf
  B915} (2017)  363--383},
\href{http://arxiv.org/abs/1608.05499}{{\tt arXiv:1608.05499 [hep-th]}}.

\bibitem{Cavaglia:2016oda}
A.~Cavagli\`{a}, S.~Negro, I.~M. Sz\'{e}cs\'{e}nyi, and R.~Tateo, \emph{{$T
  \bar{T}$-deformed 2D Quantum Field Theories}},
  \href{http://dx.doi.org/10.1007/JHEP10(2016)112}{JHEP {\bf 10} (2016)  112},
\href{http://arxiv.org/abs/1608.05534}{{\tt arXiv:1608.05534 [hep-th]}}.

\bibitem{Baggio:2018rpv}
M.~Baggio, A.~Sfondrini, G.~Tartaglino-Mazzucchelli, and H.~Walsh, \emph{{On $
  T\overline{T} $ deformations and supersymmetry}},
  \href{http://dx.doi.org/10.1007/JHEP06(2019)063}{JHEP {\bf 06} (2019)  063},
\href{http://arxiv.org/abs/1811.00533}{{\tt arXiv:1811.00533 [hep-th]}}.

\bibitem{Chang:2018dge}
C.-K. Chang, C.~Ferko, and S.~Sethi, \emph{{Supersymmetry and $ T\overline{T} $
  deformations}}, \href{http://dx.doi.org/10.1007/JHEP04(2019)131}{JHEP {\bf
  04} (2019)  131},
\href{http://arxiv.org/abs/1811.01895}{{\tt arXiv:1811.01895 [hep-th]}}.

\bibitem{Jiang:2019hux}
H.~Jiang, A.~Sfondrini, and G.~Tartaglino-Mazzucchelli, \emph{{$T\bar{T}$
  deformations with $\mathcal{N}=(0,2)$ supersymmetry}},
\href{http://arxiv.org/abs/1904.04760}{{\tt arXiv:1904.04760 [hep-th]}}.

\bibitem{Chang:2019kiu}
C.-K. Chang, C.~Ferko, S.~Sethi, A.~Sfondrini, and G.~Tartaglino-Mazzucchelli,
  \emph{{$T\bar{T}$ Flows and (2,2) Supersymmetry}},
\href{http://arxiv.org/abs/1906.00467}{{\tt arXiv:1906.00467 [hep-th]}}.

\bibitem{Coleman:2019dvf}
E.~A. Coleman, J.~Aguilera-Damia, D.~Z. Freedman, and R.~M. Soni, \emph{{$T
  \bar{T}$-Deformed Actions and (1,1) Supersymmetry}},
\href{http://arxiv.org/abs/1906.05439}{{\tt arXiv:1906.05439 [hep-th]}}.

\bibitem{Bonelli:2018kik}
G.~Bonelli, N.~Doroud, and M.~Zhu, \emph{{$T \bar{T}$-deformations in closed
  form}}, \href{http://dx.doi.org/10.1007/JHEP06(2018)149}{JHEP {\bf 06} (2018)
   149},
\href{http://arxiv.org/abs/1804.10967}{{\tt arXiv:1804.10967 [hep-th]}}.

\bibitem{Note1}
Through this Letter we follow the conventions of \cite {Farakos:2016zam}.

\bibitem{Volkov:1973ix}
D.~V. Volkov and V.~P. Akulov, \emph{{Is the Neutrino a Goldstone Particle?}},
\href{http://dx.doi.org/10.1016/0370-2693(73)90490-5}{Phys. Lett. {\bf 46B}
  (1973)  109--110}.



\bibitem{Kastor:1988ef}
  D.~A.~Kastor, E.~J.~Martinec and S.~H.~Shenker,
  \emph{RG Flow in N=1 Discrete Series},
  \href{https://www.sciencedirect.com/science/article/pii/0550321389900606}{Nucl.\ Phys.\ B {\bf 316} (1989) 590}.

\bibitem{Zamolodchikov:1991vx}
  A.~B.~Zamolodchikov,
 \emph{From tricritical Ising to critical Ising by thermodynamic Bethe ansatz},
  \href{https://www.sciencedirect.com/science/article/pii/055032139190423U}{Nucl.\ Phys.\ B {\bf 358} (1991) 524}.

\bibitem{Ferrara:1974pz}
S.~Ferrara and B.~Zumino, \emph{{Transformation Properties of the
  Supercurrent}},
\href{http://dx.doi.org/10.1016/0550-3213(75)90063-2}{Nucl. Phys. {\bf B87}
  (1975)  207}.

\bibitem{Rocek:1978nb}
M.~Ro\v{c}ek, \emph{{Linearizing the Volkov-Akulov Model}},
\href{http://dx.doi.org/10.1103/PhysRevLett.41.451}{Phys. Rev. Lett. {\bf 41}
  (1978)  451--453}.

\bibitem{Casalbuoni:1988xh}
R.~Casalbuoni, S.~De~Curtis, D.~Dominici, F.~Feruglio, and R.~Gatto,
  \emph{{Nonlinear Realization of Supersymmetry Algebra From Supersymmetric
  Constraint}},
\href{http://dx.doi.org/10.1016/0370-2693(89)90788-0}{Phys. Lett. {\bf B220}
  (1989)  569--575}.





\bibitem{Farakos:2016zam}
F.~Farakos, P.~Ko\v{c}\'{\i}, and R.~von Unge, \emph{{Superspace Higher Derivative
  Terms in Two Dimensions}},
  \href{http://dx.doi.org/10.1007/JHEP04(2017)002}{JHEP {\bf 04} (2017)  002},
\href{http://arxiv.org/abs/1612.04361}{{\tt arXiv:1612.04361 [hep-th]}}.





\bibitem{Note2}
In the limit $\lambda \to 0$, the constant term in the Volkov--Akulov action diverges. As long as we are in flat space, this (infinite) constant can be ignored.


\bibitem{Frolov:2019nrr} 
  S.~Frolov,
 \emph{{TTbar deformation and the light-cone gauge}}, 
   \href{http://arxiv.org/abs/1905.07946}{{\tt arXiv:1905.07946 [hep-th]}}.  


\bibitem{Kallosh:2014wsa} 
  R.~Kallosh and T.~Wrase, 
  \emph{{Emergence of Spontaneously Broken Supersymmetry on an Anti-D3-Brane in KKLT dS Vacua}}, 
 \href{http://dx.doi.org/10.1007/JHEP12(2014)117}{JHEP {\bf 1412} (2014) 117} 
 \href{http://arxiv.org/abs/1411.1121}{{\tt arXiv:1411.1121 [hep-th]}}. 

\bibitem{Conti:2018jho} 
  R.~Conti, L.~Iannella, S.~Negro and R.~Tateo,
  \emph{{Generalised Born-Infeld models, Lax operators and the $ \mathrm{T}\overline{\mathrm{T}} $ perturbation}},
  \href{http://dx.doi.org/10.1007/JHEP11(2018)007}{JHEP {\bf 1811} (2018) 007}
\href{http://arxiv.org/abs/1806.11515}{{\tt arXiv:1806.11515 [hep-th]}}. 

\bibitem{Taylor:2018xcy} 
  M.~Taylor,
 \emph{{TT deformations in general dimensions}}, 
  \href{http://arxiv.org/abs/1805.10287}{{\tt arXiv:1805.10287 [hep-th]}}.  






\end{thebibliography}
\end{document}